\documentclass[apl,twocolumn]{revtex4-1}

\usepackage{graphicx}
\usepackage{color}
\begin{document}

\title{Atomically Sharp 318nm Gd:AlGaN Ultraviolet Light Emitting Diodes on Si with Low Threshold Voltage}

\author{Thomas F. Kent}
\affiliation{Department of Materials Science and Engineering, The Ohio State University, Columbus, Ohio 43210, USA}

\author{Santino D. Carnevale}
\affiliation{Department of Materials Science and Engineering, The Ohio State University, Columbus, Ohio 43210, USA}

\author{Roberto C. Myers}
\affiliation{Department of Materials Science and Engineering, The Ohio State University, Columbus, Ohio 43210, USA}
\affiliation{Deparment of Electrical and Computer Engineering, The Ohio State University, Columbus, Ohio 43210, USA}

\begin{abstract}
Self-assembled Al$_x$Ga$_{1-x}$N polarization-induced nanowire light emitting diodes (PINLEDs) with Gd-doped AlN active regions are prepared by plasma-assisted molecular beam epitaxy on Si substrates. Atomically sharp electroluminescence (EL) from Gd intra-f-shell electronic transitions at 313nm and 318nm is observed under forward biases above 5V. The intensity of the Gd 4f EL scales linearly with current density and increases at lower temperature. The low field excitation of Gd 4f EL in PINLEDs is contrasted with high field excitation in metal/Gd:AlN/polarization-induced n-AlGaN devices; PINLED devices offer over a three fold enhancement in 4f EL intensity at a given device bias.
\end{abstract}

\maketitle

In this letter we report on ultraviolet light emitting diodes based on self-assembled AlGaN nanowire heterojunctions doped with Gd, which emit ultraviolet (UV) radiation at 318nm and operate at lower device bias compared with existing Gd electroluminescent technology. Following the realization of the ruby laser over half a century ago\cite{maiman}, many materials have been developed as hosts for elements with optically-active energetic transitions to produce lasers and electroluminescent devices with emission in the visible and infrared part of the electromagnetic spectrum. In the last twenty years, electrically driven devices utilizing rare earth phosphors in wide or medium gap semiconductors have begun to make an appearance\cite{rack}. These thin film electroluminescent devices (TFED) have found applications as visible and IR emitters, and with the increasing use of wide gap materials, such as GaN and AlN are currently being developed to take of advantage of phosphors in the blue and UV.

\begin{figure}[h]
\center
\includegraphics[scale=0.875]{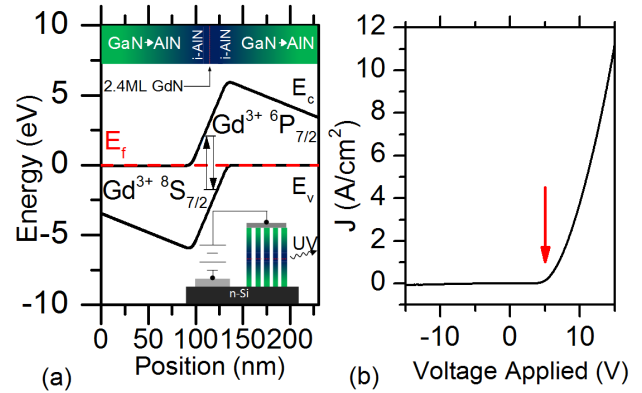}
\caption{Optically active Gd$^{3+}$ 4f levels in the active region of a III-N nanowire polarization induced light emitting diode. \textbf{a.} Calculated device band diagram showing location of optically active Gd ions. Insert shows device schematic as measured. \textbf{b.} Device IV exhibiting turn on in forward bias at 5V}
\label{bipolardev}
\end{figure}

\begin{figure*}
\center
\includegraphics[scale=0.9]{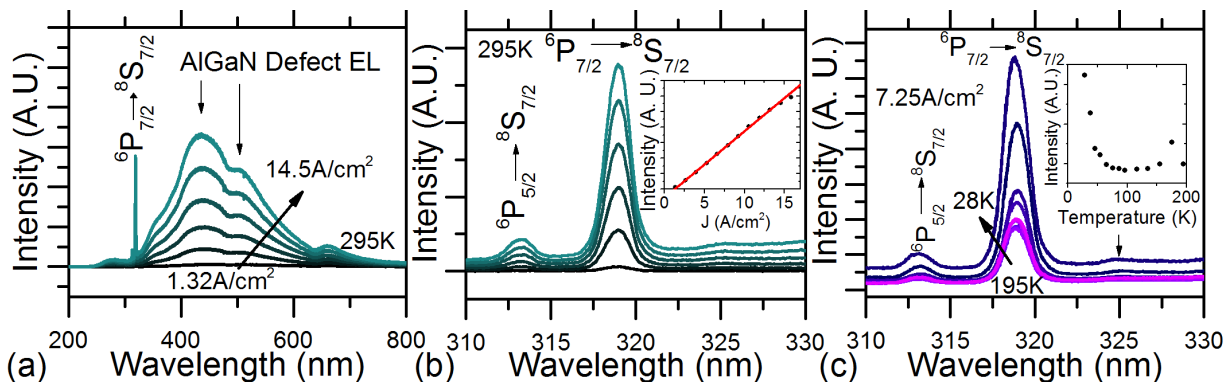}
\caption{Optically active Gd ions in the depletion region of a polarization-induced light emitting diode (PINLED). \textbf{a.} Room temperature device EL spectra for multiple currents. \textbf{b.} Detail of spectral region of interest for the Gd 4f shell EL peaks, showing first excited and second excited state transitions. Insert shows behavior of peak intensity with increasing current density. \textbf{c.} Detail of spectra region of interest for the Gd 4f EL at multiple temperatures. Insert shows variation of peak intensity with temperature.}
\label{bipolar}
\end{figure*}
Design of optoelectronic devices which utilize atomic transitions in the rare earths for EL can offer a number of advantages over EL produced from band-to-band recombination. First, the transition energy is dictated by the energy level scheme of the \textit{4f} orbital, which is relatively unperturbed by the crystalline environment due to the fact that the \textit{5d} and \textit{5s,5p} orbitals extend to further radial distances, and fill before the \textit{f} shell of lower principal quantum number\cite{atkins}. This shielding by the earlier filled \textit{5d} and \textit{5s} orbitals causes the energies of the \textit{4f} transitions to be relatively insensitive to crystalline imperfections, unlike transitions based on band-to-band transitions in semiconductors. Band-to-band optical transitions are well known\cite{davies02} to be sensitive to deep levels, exciton-phonon interaction, and crystalline disorder which can lead to broadening of emission or parasitic emission at an unintended energy. Additionally, due to the decoupling of the 4f orbital with the lattice, emission from rare earth centers is spectrally pure with common FWHM of less than 30meV.

In recent years, the family of III-nitrides, particularly the pseudobinary Al$_{x}$Ga$_{1-x}$N, has become attractive as a potential host for rare earth elemental phosphors\cite{wakahara}. This material system offers a high breakdown field (12MV/cm for AlN) and sufficiently wide bandgaps (3.4 to 6.1eV) to accommodate phosphors with emission from the UV to IR. Equally significant is the way in which rare earths incorporate in the wurtzite structure of GaN and AlN alloys. Rare earth (RE) ions commonly exhibit the RE$^{3+}$ ionization state, which makes them isovalent with the Al$^{3+}$ and Ga$^{3+}$ cations of Al$_{x}$Ga$_{1-x}$N. RE atoms exhibit high solubility (in excess of 1at\% has been reported\cite{wang}) and regularly incorporate at the cation site.  Due to the wide range of energy level schemes of the lanthanides, EL from RE centers in AlN and GaN have been reported for Er$^{3+}$ (IR, green)\cite{Lee}, Tm$^{3+}$(blue)\cite{Lee}, Eu$^{3+}$(Red)\cite{wakahara}, and Gd$^{3+}$(UV)\cite{kita}. Additonally, room temperature optically pumped lasing in Eu-doped GaN\cite{park} has been successfully demonstrated.

	Although most RE phosphors have been developed for optical transitions in the visible and infrared parts of the electromagnetic spectrum, the energy scheme of Gd$^{3+}$ in wurtzite AlN offers an energy difference between ground and first excited state  of 3.90eV (318nm). The spectrally-narrow and energetically-stable nature of the Gd$^{3+}$ fluorescence emission make it a potential candidate for spectroscopic and lithographic applications in the UV. This led to exploration of dilutely Gd doped Al$_x$Ga$_{1-x}$N in the form of fluorescence\cite{zavada} and cathodoluminescence experiments\cite{vetter, gruber, kita}.Although the 4f levels in the RE$^{3+}$ are typically thought not to interact with the surrounding lattice, cathodoluminescence data for Gd:AlN thin films show phonon replica satellite peaks of the Gd$^{3+}$ $^6$P$_{7/2}$$\rightarrow$$^{8}$S$_{7/2}$  (318nm) transitions\cite{vetter}. These data suggest that the \textit{4f} electrons in Gd$^{3+}$ in AlN are not completely decoupled from the host lattice.

Although there have been a number of reports\cite{kita,zavada,vetter,gruber} on the spectroscopy of Gd:AlGaN compounds, less work has focused on development of active optoelectronic devices that utilize Gd$^{3+}$ \textit{4f} transitions. This is likely due to the difficulty of achieving electrical contact to uid-AlN.  Reports\cite{kita, kitayama} have been made of a ``field emission device"  consisting of a reactive ion sputtered Al$_x$Gd$_{1-x}$N film with metal contacts, forming a MIS structure whereby a high voltage 270V to $>$1kV driven across the device produces fluorescence of the Gd$^{3+}$ ions, likely by the process of impact excitation.
\begin{figure*}
\center
\includegraphics[scale=0.9]{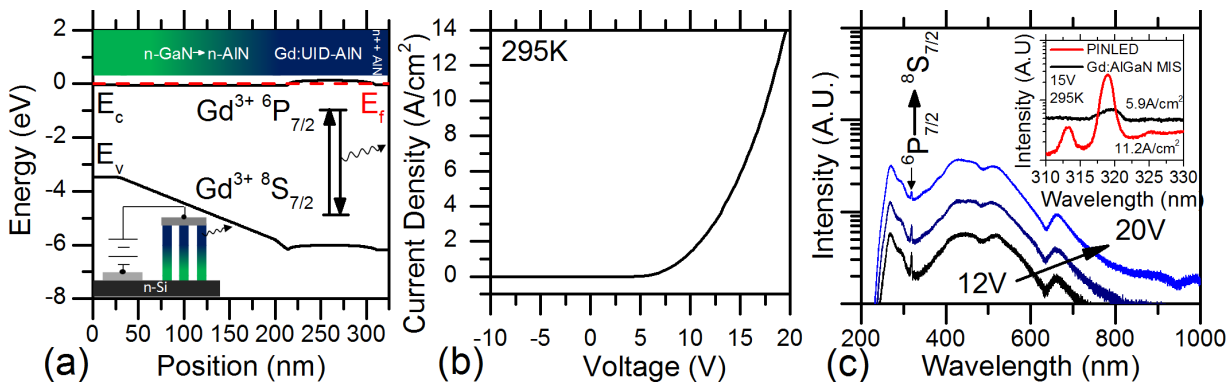}
\caption{Optically active Gd ions in a purely unipolar metal-insulator-semiconductor heterostructure \textbf{a.} Device heterostructure and calculated band diagram. Insert shows device schematic. \textbf{b.} Device IV curve exhibiting rectification and a turn on voltage above 10V. \textbf{c.} EL spectra showing weak EL of the Gd 4f $^6$P$_{7/2}$$\rightarrow$$^{8}$S$_{7/2}$ transition. Insert shows detail of the 4f EL peak with an overlay of the more intense spectrum from the PINLED device under comparable conditions.}
\label{unipolar}
\end{figure*} 

The dominant mechanism of RE$^{3+}$ 4f excitation in solids depends on the electric field regime where the device functions. MIS devices function largely in the high field regime, causing the dominant mechanism to be direct transfer of kinetic energy from hot electrons to the RE$^{3+}$ center by collision, known as impact excitation. At lower fields, the excitation mechanism becomes more complex, involving typically a multi-step, defect assisted Auger process, or exciton localization by the RE$^{3+}$ center.\cite{godlewski,bodiou} In this work, we study two seperate heterostructures which are designed to generate EL under low electric field conditions (pn-diode) and high electric field conditions (Gd:AlGaN MIS structure). In the PINLED, the initially large built in electric field in the depletion region is reduced with forward bias of the device. In contrast, the MIS structure has initially flat bands and thus no electric field in the active region and when biased, the electric field is increased in the Gd doped region. 

Self-assembled III-nitride nanowire heterostructures grown by plasma assisted molecular beam epitaxy have recently gained popularity for applications requiring high crystalline quality and largely mismatched, complex heterostructures that would otherwise be difficult to form in thin films due to strain considerations\cite{carnevale1,carnevale2,carnevale3}. In addition, they have been shown to function at high current densities\cite{carnevale3}. Additionally, UV-LEDs based on III-nitride nanowire heterostructures have been demonstrated to accommodate active regions spanning from high \%Al AlGaN to GaN\cite{carnevale3}.

Polarization-induced nanowire diodes (PINLEDs)\cite{carnevale2} containing Gd doping in their active region are prepared by plasma-assisted molecular beam epitaxy on n-Si(111) in the III-limited growth regime, the details of which are discussed elsewhere\cite{carnevale2}. These devices consist of a GaN nucleation layer followed by a linear grade in composition from GaN to AlN over 100nm. This is followed by an active region consisting of 2.4ML of GdN deposited between two 5nm uid-AlN spacers. The structure is then linearly graded in composition from AlN back to GaN. This structure forms a pn-diode as shown in Fig.~\ref{bipolardev}a, the band diagram of which is calculated with a self consistent Schr\"{o}dinger-Poisson solver\cite{bandeng}, as described in ref\cite{carnevale2}. The spontaneous polarization present in wurtzite Al$_{x}$Ga$_{1-x}$N combined with a gradient in composition give rise to polarization-induced hole and electron doping, respectively\cite{simon}. In addition to the built in polarization doping, wires are supplementary doped with Mg and Si in p and n regions, assuming c-axis,Ga-polar orientation of the nanowires. 

From these heterostructures, electrical devices are fashioned by depositing semitransparent 10nm/20nm Ni/Au contacts with an electron beam evaporator for the top contact, which connect the vertical ensemble in parallel. The back contact is fashioned by mechanically removing the nanowires adjacent to the top contact with a diamond scribe and thermally diffusing In metal directly to the n-Si with a soldering ion. Current-Voltage (IV) behavior of the devices are measured with a probe station and an Agilent B1500 semiconductor parameter analyzer. Device IV's show rectifying behavior with device turn on at 5V, as shown in Fig.~\ref{bipolardev}b.

After the IV behavior of the devices is characterized, they are transferred to a variable temperature UV-VIS-NIR spectroscopy system consisting of a closed cycle ARS DMX20-OM cryostat and a Princeton instruments SP2500i 0.5m spectrometer equipped with a Princeton instruments PIXIS100 UV-VIS-NIR CCD. Devices are connected to a Yokogawa DC constant current source and Keithley 2700 data acquisition system. Prior to the collection of any spectra, a background spectrum is collected with no current injection in the device. Constant currents from 10mA to 120mA, corresponding to device current densities from 1.32A/cm$^2$ to 14.5A/cm$^2$ (the current density through any given nanowire is unknown since not all individual nanowires give EL\cite{limbach}) are sourced at room temperature and the resulting EL is collected through a 50mm, f/2 uv-fused silica singlet lens, collimated and subsequently focused onto the entrance slit of the 0.5m spectrometer. 

Room temperature EL spectra, shown for multiple current densities in Fig.~\ref{bipolar}a., exhibit multiple emission peaks from the UV through the visible parts of the spectrum. The sharp peak at 318nm (Fig.~\ref{bipolar}b.), corresponds to the $^6$P$_{7/2}$$\rightarrow$$^{8}$S$_{7/2}$ first excited state to ground state transition of the Gd$^{3+}$ ion\cite{gruber}. More careful inspection of spectra around the Gd atomic line region reveals an additional peak at the correct energy for the $^6$P$_{5/2}$$\rightarrow$$^{8}$S$_{7/2}$ second excited state to ground state transition\cite{gruber}. In addition, the intensity of these peaks scale linearly with current density, within the range of currents investigated. The Gd emission exhibits FWHM of 23.1meV. Additional broad EL peaks in the 400nm-700nm range are attributed to below band gap defects in AlGaN, due to observation of identical emission spectra in non-Gd containing devices, though the graded nature of structure prevents precise identification of the deep levels responsible.

In order to further investigate the mechanism by which the UV emission occurs, variable temperature EL measurements are conducted from room temperature to 30K, the results of which are shown in Fig.~\ref{bipolar}c. Spectral intensity of the main $^6$P$_{7/2}$$\rightarrow$$^{8}$S$_{7/2}$ Gd 4f peak is observed to be invariant at temperatures above 75K, below which the intensity increases dramatically (Fig.~\ref{bipolar}c., insert). Similar behavior has been previously observed in Eu-doped GaN\cite{nyein} as well as Er-doped InGaP\cite{neuhalfen} which is attributed to thermal quenching of the multi-step excitation mechanism of the RE$^{3+}$ ion. At low temperature another peak becomes distinct at 324nm. This peak has been previously identified as a phonon replica of the primary 318nm peak in cathodoluminescence experiments\cite{vetter}. From analysis of the peak positions, a phonon mode energy of 72.2meV is measured, which is smaller than the LO phonon energy of the surrounding AlN matrix (110meV\cite{bergman}) as well as GaN (92meV\cite{cingolani}). This phonon energy agrees within the resolution of the spectrometer used with results from cathodoluminescence experiments, which report 72.9meV\cite{vetter}. 

 Many RE electroluminescent devices rely on the impact excitation mechanism to drive intra-f-shell EL. In the interest of investigating EL under these conditions in nanowire based devices, a structure consisting of an n-type nanowire graded from GaN to AlN, with a 200nm uid-AlN layer doped with Gd ions (1E18cm$^{-3}$) and capped with a small amount of n++ AlN for a top contact is prepared. Growth conditions for the heterostructure are identical to the n-region and depletion region of the PINLED, thus similar defect content are expected in both structures. An identical device contact scheme to that of the heterojunction diode is used and is shown in Fig.~\ref{unipolar}a. This device is again rectifying, as shown in Fig.~\ref{unipolar}b., but is less conductive than the PINLED device, producing 5.9A/cm$^2$ compared to 11.2A/cm$^2$ when forward biased to 15V. At 15V it is calculated that the active region of the MIS device develops an electric field in excess of 0.75MV/cm, where as the PINLED should have approximately flat band conditions in the active region, due to reduction of band bending in forward bias. The lower conductivity of the MIS devices can be attributed to the uid-AlN center region as well as a large Schottky barrier between the n++AlN and the Ti/Au top contact which produce additional series resistance in the device over the PINLED. Electroluminescence spectroscopy (Fig.~\ref{unipolar}c.) reveals a weak, but detectable peak at 318nm among a large background of defect luminescence, indicating that some Gd ions are being excited by hot electrons passing through the structure as well as impact excitation of band to defect luminescence. Comparison of emission from PINLED devices and Gd:AlGaN MIS structures (Fig.~\ref{unipolar}c., insert) shows a 372\% enhancement of intensity of the $^6$P$_{7/2}$$\rightarrow$$^{8}$S$_{7/2}$ transition in the PINLED devices over the Gd:AlGaN MIS devices at 15V bias. Additionally, no peak corresponding to the $^6$P$_{5/2}$$\rightarrow$$^{8}$S$_{7/2}$ higher order transition is present in the emission spectra from the Gd:AlGaN MIS structure. It is noted that due to the variety of possible mechanisms for RE 4f excitation in III-V materials, the difference in performance between the MIS device and the PINLED device could be affected by phenomena which are extrinsic to the E-field regime, such as preferential interaction of one carrier type with the RE ion. 

In conclusion, polarization induced light emitting diodes (PINLEDS) doped with Gd in an AlN active region are prepared by plasma assisted molecular beam epitaxy on n-Si substrates. These devices function at one to two orders of magnitude lower biases than previously reported Gd:AlN electroluminescent devices, making them attractive for low power, UV EL applications, particularly portable devices. When forward biased, devices emit sharp peaks at 318nm and  313nm, which correspond to the Gd intra-f-shell $^6$P$_{7/2}$$\rightarrow$$^{8}$S$_{7/2}$ and  $^6$P$_{5/2}$$\rightarrow$$^{8}$S$_{7/2}$ transitions, respectively and scale linearly with current density. Emission intensity is shown to be temperature independent above 75K, below which it increases strongly. By studying two different devices, designed to produced Gd 4f EL under both low and high electric field conditions, we observe a significant improvement in emission intensity for PINLED devices which function under low-field operation conditions over hot electron MIS devices. Although this device has been applied to Gd, it would be possible in principle to dope with any of the 4f phosphor rare earths to achieve spectrally stable electrically driven emission at a variety of wavelengths.

	This work is supported by the Center of Emergent Materials at OSU under NSF DMR-0820414 and National Science Foundation CAREER award (DMR-1055164). S. D. Carnevale acknowledges support from the National Science Foundation Graduate Research Fellowship Program (2011101708).

\end{document}